\documentclass[aip,graphicx]{revtex4-1}

\usepackage{graphicx}
\DeclareGraphicsExtensions{.pdf,.png,.jpg,.eps}

\draft 

\begin{document}


\title{Representation Independent Boundary Conditions for a Piecewise-Homogeneous Linear Magneto-dielectric Medium} 



\author{Michael E. Crenshaw}
\affiliation{Charles M. Bowden Research Laboratory, US Army Combat Capabilities Development Command (DEVCOM) - Aviation and Missile Center, Redstone Arsenal, AL 35898, USA}


\date{\today}

\begin{abstract}
At a boundary between two transparent, linear, isotropic, homogeneous
materials, derivations of the electromagnetic boundary conditions and the
Fresnel relations typically proceed from the
Minkowski \{{\bf E},{\bf B},{\bf D},{\bf H}\}
representation of the macroscopic Maxwell equations.
However, equations of motion for macroscopic fields in a transparent linear
medium can be written using Amp\`ere \{{\bf E},{\bf B}\},
Chu \{{\bf E},{\bf H}\}, Lorentz, Minkowski, Peierls, Einstein--Laub, and
other formulations of continuum electrodynamics.
We present a representation-independent derivation of electromagnetic
boundary conditions and Fresnel relations for the propagation of
monochromatic radiation through a piecewise-homogeneous, transparent,
linear, magneto-dielectric medium.
The electromagnetic boundary conditions and the Fresnel relations are
derived from energy conservation coupled with the application of Stokes's
theorem to the wave equation.
Our representation-independent formalism guarantees the general
applicability of the Fresnel relations.
Specifically, the new derivation is necessary so that a valid derivation
of the Fresnel equations exists for alternative, non-Minkowski
formulations of the macroscopic Maxwell field equations.
\end{abstract}


\maketitle 

\par
In continuum dynamics, the usual procedure to derive boundary conditions
is to apply conservation of energy and conservation of momentum across
the surface of a control volume that contains the boundary. \cite{BIFox}
However, such a procedure fails in continuum electrodynamics because the
energy and momentum of light are not linearly independent as both depend
quadratically on the macroscopic fields.
For this case, rather than use conservation principles, the 
electromagnetic boundary conditions and the Fresnel relations are
typically obtained by the application of Stokes's theorem and the
divergence theorem to the Maxwell--Minkowski
equations \cite{BIMarion,BIGriffiths,BIJack,BIZang}
\begin{equation}
\nabla\times{\bf E}+\frac{1}{c}\frac{\partial{\bf B}}{\partial t}=0
\label{EQp1.01a}
\end{equation}
\begin{equation}
\nabla\times{\bf H}-\frac{1}{c}\frac{\partial{\bf D}}{\partial t}=
\frac{{\bf J}_f}{c}
\label{EQp1.01b}
\end{equation}
\begin{equation}
\nabla\cdot{\bf B}=0 
\label{EQp1.01c}
\end{equation}
\begin{equation}
\nabla\cdot{\bf D}= \rho_f+\rho_b\, .
\label{EQp1.01d}
\end{equation}
The macroscopic Minkowski fields,
${\bf E}({\bf r},t)$, ${\bf D}({\bf r},t)$,
${\bf B}({\bf r},t)$, and ${\bf H}({\bf r},t)$,
are functions of position ${\bf r}$ and time $t$.
Here, ${\bf J}_f({\bf r},t)$ is the free current,
$\rho_f({\bf r},t)$ is the free charge density,
and $\rho_b({\bf r})$ is the bound charge density.
The macroscopic electric and magnetic fields are related by the 
constitutive relations
\begin{equation}
{\bf D}=\varepsilon{\bf E}
\label{EQp1.02a}
\end{equation}
\begin{equation}
{\bf B}=\mu{\bf H} \, ,
\label{EQp1.02b}
\end{equation}
where $\varepsilon({\bf r},t)$ is the electric permittivity and
$\mu({\bf r},t)$ is the magnetic permeability.
\par
As equations of motion for macroscopic electromagnetic fields in matter,
the Maxwell--Minkowski equations are not unique.
Alternative formulations that are associated with Amp\`ere, 
Chu, Lorentz, Minkowski, Peierls, Einstein--Laub, and
and others \cite{BIPenHaus,BIKemp,BIKempRev,BIPeierls,BIJMP,BIMol}
are sometimes used to emphasize various features of classical
electrodynamics in matter.
In the Chu formalism of
electrodynamics, \cite{BIPenHaus,BIKemp,BIKempRev} for example,
\begin{equation}
\nabla\times{\bf E}^c+\frac{1}{c}\frac{\partial{\bf H}^c}{\partial t}=
-\frac{1}{c}\frac{\partial{\bf M}^c}{\partial t}
-\nabla\times\left ({\bf M}^c\times{\bf v} \right )
\label{EQp1.03a}
\end{equation}
\begin{equation}
\nabla\times{\bf H}^c-\frac{1}{c}\frac{\partial{\bf E}^c}{\partial t}=
\frac{1}{c}\frac{\partial{\bf P}^c}{\partial t}
+\nabla\times \left ({\bf P}^c \times{\bf v} \right ) 
+\frac{{\bf J}^c}{c}
\label{EQp1.03b}
\end{equation}
\begin{equation}
\nabla\cdot{\bf H}^c=-\nabla\cdot{\bf M}^c
\label{EQp1.03c}
\end{equation}
\begin{equation}
\nabla\cdot{\bf E}^c= -\nabla\cdot{\bf P}^c +\rho^c \, ,
\label{EQp1.03d}
\end{equation}
the material response is separated from the Chu electric field ${\bf E}^c$
and the Chu magnetic field ${\bf H}^c$. The Chu polarization ${\bf P}^c$
and Chu magnetization ${\bf M}^c$ are treated as sources for the fields.
\par
Although most of the different expressions of macroscopic electromagnetic
theory appear to be justified, there have also been some questions about
validity.
Before the work of Penfield and Haus, \cite{BIPenHaus} it was believed
that optical forces predicted by the Chu formulation,
Eqs.~(\ref{EQp1.03a})-- Eqs.~(\ref{EQp1.03d}),
differed from other accepted theories. \cite{BIKemp,BIKempRev}
Also, the validity of the Einstein--Laub formulation is still debated.
\cite{BIMasud,BIShepKemp}
The problem that we address is that, in some of these
representations, \cite{BIJMP} the Fresnel relations cannot be derived
by the straightforward application of textbook techniques to the field
equations.
A general derivation is needed to avoid arguments that boundary conditions
are violated in some representations based on the inapplicability of a
technique that was developed for the Minkowski representation of the
macroscopic Maxwell field equations.
\par
In this article, the electromagnetic boundary conditions and the Fresnel
relations are derived for piecewise-homogeneous transparent linear
magneto-dielectric media from conservation of energy coupled with an
application of Stokes's theorem to the wave equation.
The new derivation is necessary so that a valid derivation of the
Fresnel relations exists for alternative, non-Minkowski
formulations of the macroscopic Maxwell field
equations. \cite{BIPenHaus,BIKemp,BIKempRev,BIPeierls,BIJMP,BIMol}
As long as energy is conserved and the wave equation is valid in a 
specific formulation of continuum electrodynamics for a transparent
linear medium, the Fresnel relations will also be valid.
\par
We consider an arbitrarily long, nominally monochromatic pulse of light
in the plane-wave limit to be propagating through a transparent, linear,
homogeneous, magneto-dielectric medium with permittivity $\varepsilon_1$
and permeability $\mu_1$ to be incident on a plane interface with
a second homogeneous transparent linear medium with permittivity
$\varepsilon_2$ and permeability $\mu_2$.
The frequency of the field is assumed to be sufficiently far from
material resonances that absorption can be neglected.
The complete system that consists of a simple transparent linear
magneto-dielectric medium plus finite radiation field isolated in free
space is thermodynamically closed.
\par
Acceleration of the material due to optically induced forces is viewed as
negligible in the lowest order theory.
The condition of a stationary, piecewise-homogeneous, linear medium
with negligible absorption illuminated by an arbitrarily long, nominally
monochromatic field (square/rectangular shape or ``top-hat'' finite
pulse) in the plane-wave limit is a physical abstraction that is used in
textbook derivations of the Fresnel relations from the Maxwell--Minkowski
equations. \cite{BIMarion,BIGriffiths,BIJack,BIZang}
We use the same conditions in our derivation.
Although ``real-world'' materials are more complicated than the theoretical 
models, appeal to complexity does not invalidate the theory for the
idealized model.
\par
For a transparent, linear, magneto-dielectric material, the energy formula
depends on the representation that is used for the macroscopic fields.
In the usual Maxwell--Minkowski representation, the total electromagnetic
energy 
\begin{equation}
U=\int_{\sigma}\frac{1}{2}
\left ({\bf E}\cdot{\bf D}+{\bf B}\cdot{\bf H} \right ) dv  \, ,
\label{EQp1.04}
\end{equation}
is conserved in a volume of space that contains all fields present
by extending the region of integration to all-space $\sigma$.
For a finite homogeneous medium, we can write the energy formula as
\begin{equation}
U(V)=\int_{V}\frac{1}{2}\left ( 
\left ( \frac{n_e}{c} \frac{\partial{\bf A}}{\partial t}\right )^2
+\left ( \frac{ \nabla\times{\bf A}}{n_m} \right )^2
\right ) dv  \, ,
\label{EQp1.05}
\end{equation}
where $V$ is the integration volume and
\begin{equation}
{\bf E}({\bf r},t)=-\frac{1}{c}\frac{\partial{\bf A}({\bf r},t)}{\partial t}
\label{EQp1.06a}
\end{equation}
\begin{equation}
{\bf B}({\bf r},t)=\nabla\times{\bf A}({\bf r},t) \, .
\label{EQp1.06b}
\end{equation}
The electric refractive index $n_e$ and the magnetic refractive
index $n_m$ can be related to the familiar permittivity,
$\varepsilon=n_e^2$, and permeability, $\mu=n_m^2$, in the
Maxwell--Minkowski formulation of continuum electrodynamics for
a simple linear medium.
The use of $n_e$ and $n_m$ instead of $\sqrt{\varepsilon}$ and $\sqrt{\mu}$,
respectively, is a matter of notational convenience. 
\par
For linearly polarized radiation normally incident on the boundary
between two homogeneous linear media, we can write the vector
potential of the incident ($i$), reflected ($r$), and
refracted ($R$) waves in terms of the constant amplitudes
$\tilde A_i$, $\tilde A_r$, and $\tilde A_R$ of the rectangular/top-hat
fields as \cite{BIMarion,BIGriffiths,BIJack,BIZang}
\begin{equation}
{\bf A}_i={\bf\hat e}_x\tilde A_i e^{-i(\omega_d t-{k}_1 z)}
\label{EQp1.07a}
\end{equation}
\begin{equation}
{\bf A}_r={\bf\hat e}_x\tilde A_r e^{-i(\omega_d t+{k}_1 z)}
\label{EQp1.07b}
\end{equation}
\begin{equation}
{\bf A}_R={\bf\hat e}_x\tilde A_R e^{-i(\omega_d t-{k}_2 z)}
\label{EQp1.07c}
\end{equation}
in the plane-wave limit.
Here, $\omega_d$ is the frequency of the field propagating
in the direction of the $z$-axis and ${\bf\hat e}_x$ is a unit
polarization vector that is perpendicular to the direction of
propagation.
\par
The total energy of a thermodynamically closed system is constant in
time by virtue of being conserved.
At time $t_0$, the entire field is in medium 1 propagating toward
the interface with medium 2.
At time $t_1$, the entire refracted field is in medium 2 and the
reflected field is in medium 1. Both the refracted and reflected
fields are propagating away from the material interface.
The total energy at time $t_0$, the incident energy $U(t_0)=U_i$,
is equal to the total energy at a later time $t_1$, $U(t_1)=U_r+U_R$,
which is the sum of the reflected energy $U_r(t_1)$ and the refracted
energy $U_R(t_1)$.
In terms of the incident, reflected, and refracted energy, the energy
balance $U_{total}(t_0)=U_{total}(t_1)$ is 
\begin{equation}
U_i(t_0)=U_r(t_1)+U_R(t_1) \, .
\label{EQp1.08}
\end{equation}
Substituting Eqs.~(\ref{EQp1.07a})--(\ref{EQp1.07c})
into the formula
for the energy, Eq.~(\ref{EQp1.05}), and expressing the energy 
balance, Eq.~(\ref{EQp1.08}), in
terms of the amplitudes of the incident, reflected, and refracted
vector potentials results in
\begin{equation}
\int_{V_1} {{n_e}_1^2} \tilde A_i^2 dv=
\int_{V_1} {{n_e}_1^2} \tilde A_r^2 dv+
\int_{V_2} {{n_e}_2^2} \tilde A_R^2 dv 
\label{EQp1.09}
\end{equation}
upon cancelling common constant factors.
Here, $V_1$ is the spatial volume of the incident/reflected field
in the vacuum and $V_2$ is the volume of the refracted field in the
medium.
In order to facilitate the integration of Eq.~(\ref{EQp1.09}), we
choose the incident pulse to be rectangular with a nominal width of
$w_i$.
The refracted pulse in medium 2 has a width of
${n_e}_1 {n_m}_1w_i/({n_e}_2{n_m}_2)$
due to the change in the velocity of light between the two media.
Then, evaluating the integrals of Eq.~(\ref{EQp1.09}) results in
\begin{equation}
\frac{{n_e}_1}{{n_m}_1} \tilde A_i^2 w_i
=\frac{{n_e}_1}{{n_m}_1} \tilde A_r^2 w_i
+\frac{{n_e}_2}{{n_m}_2} \tilde A_R^2  w_i\,.
\label{EQp1.10}
\end{equation}
Grouping terms of like refractive index, the previous equation becomes
\begin{equation}
\frac{{n_e}_1}{{n_m}_1} \left (\tilde A_i^2 - \tilde A_r^2 \right ) =
\frac{{n_e}_2}{{n_m}_2} \tilde A_R^2  \, .
\label{EQp1.11}
\end{equation}
The second-order equation, Eq.~(\ref{EQp1.11}), can be written as two
first-order equations, but the decomposition is not unique.
\par
In order to derive boundary relations, we need a second linearly
independent relation.
Substituting the relations between the vector potential and the fields,
Eqs.~(\ref{EQp1.06a}) and (\ref{EQp1.06b}), into the Maxwell--Amp\`ere
law, Eq.~(\ref{EQp1.01b}), we obtain
the wave equation
\begin{equation}
\nabla\times\frac{\nabla\times{\bf A}({\bf r},t)}{(n_m({\bf r}))^2}+
\frac{(n_e({\bf r}))^2}{c^2}
\frac{\partial^2{\bf A}({\bf r},t)}{\partial t^2}=0
\label{EQf1.08}
\end{equation}
that describes the propagation of the electromagnetic field through an
optically transparent linear magneto-dielectric medium.
\par
Consider a thin right rectangular box or ``Gaussian pillbox'' that
straddles the interface between the two mediums with the large surfaces
parallel to the interface, Fig.~1.
Then $S$ is the surface of the pillbox,
$da$ is an element of area on the surface, and ${\bf\hat n}$
is an outwardly directed unit vector that is normal to $da$.
Integrating the wave equation, Eq.~(\ref{EQf1.08}), we obtain
$$
\int_S\left (
\nabla\times \frac{\nabla\times{\bf A}({\bf r},t)}
{(n_m({\bf r}))^2}
\right )
\cdot{\bf\hat n} \, da =
$$
\begin{equation}
-\int_S\frac{(n_e({\bf r}))^2}{c^2}
\frac{\partial^2{\bf A}({\bf r},t)}{\partial t^2}\cdot
{\bf\hat n} \, da \, .
\label{EQf1.09}
\end{equation}
There is no contribution to the surface integral from the large surfaces
for a normally incident field in the plane-wave limit because both
${\bf A}$ and $\nabla\times(\nabla\times{\bf A})$ are orthogonal to
${\bf\hat n}$.
The contributions from the smaller surfaces can be neglected as the box
becomes arbitrarily thin. 
\begin{figure}
\includegraphics[]{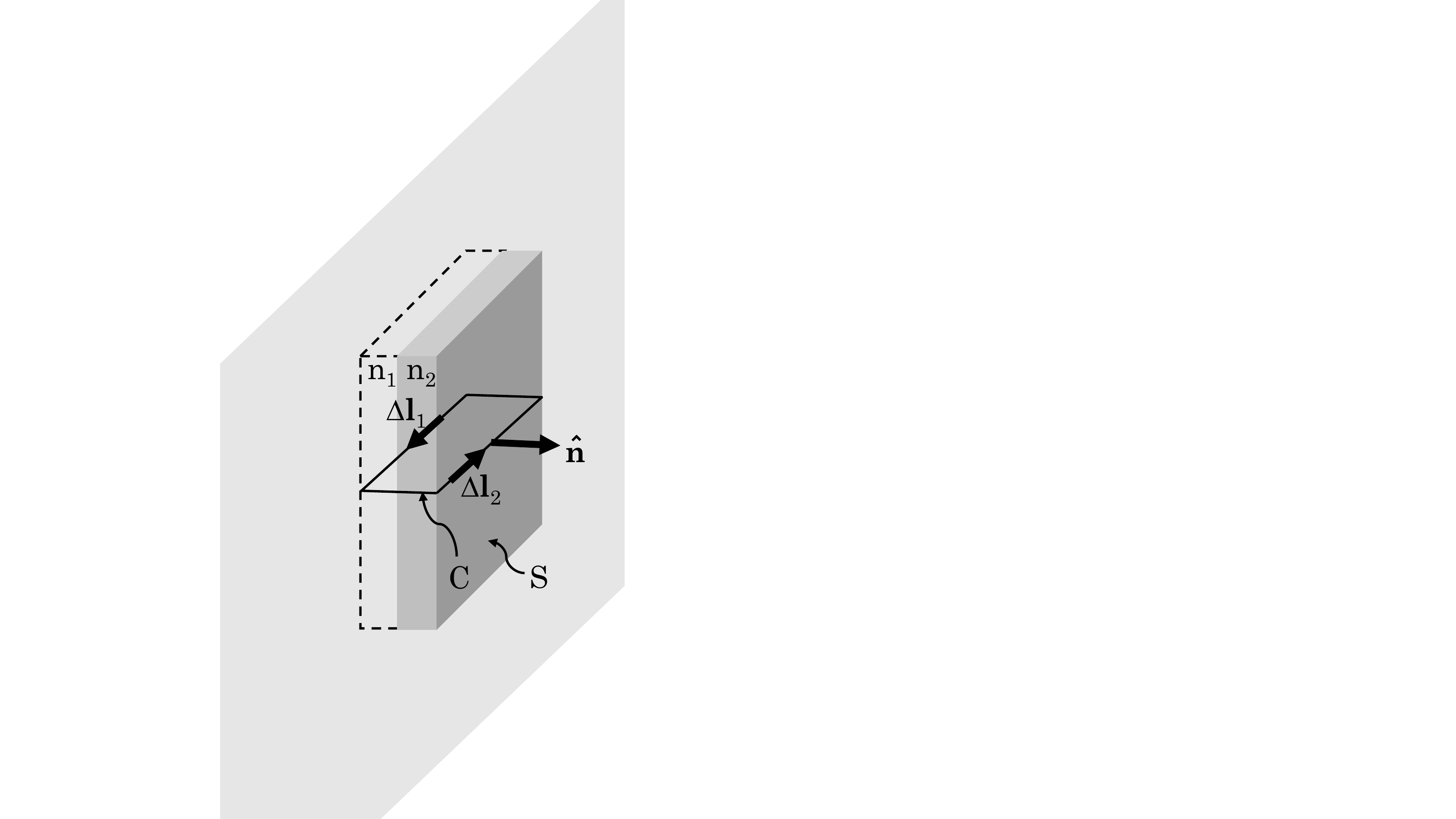}
\caption{An interface between two magneto-dielectric materials,
$\{
{\varepsilon}_1,{\mu}_1
\}$
and
$\{
{\varepsilon}_2,{\mu}_2
\}$,
illustrating the ``Gaussian pillbox'' and the Stokesian loop across
the interface.}
\end{figure}
Applying Stokes's theorem for an arbitrary vector ${\bf v}$
\begin{equation}
\oint_C {\bf v} \cdot d{\bf l}=
\int_S (\nabla\times{\bf v})\cdot{\bf\hat n} \, da 
\label{EQf1.10}
\end{equation}
to Eq.~(\ref{EQf1.09}),
we have
\begin{equation}
\oint_C\frac{\nabla\times{\bf A}}{(n_m({\bf r}))^2}\cdot d{\bf l}= 0\, .
\label{EQf1.11}
\end{equation}
We choose the closed contour $C$ in the form of a rectangular Stokesian
loop with sides that bisect the two large surfaces and two of the small
surfaces on opposite sides of the pillbox shown in Fig.~1.
Here, $d{\bf l}$ is a directed line element that lies on the
contour, $C$.
Then $C$ straddles the material interface.
For normal incidence in the plane-wave limit, the field
$n_m^{-2}\nabla\times{\bf A}$ can be oriented along the long sides of
the contour $C$.
Performing the contour integration in Eq.~(\ref{EQf1.11}), the
contribution from the short sides of the contour are neglected as the
loop is made vanishingly thin and we obtain
\begin{equation}
\frac{1}{({n_m}_1)^2}
(\nabla\times{\bf A})_1\cdot\Delta {\bf l}_1 +
\frac{1}{({n_m}_2)^2}
(\nabla\times{\bf A})_2\cdot\Delta {\bf l}_2=0
\label{EQf1.12}
\end{equation}
from the long sides, 1 and 2, of the contour.
\par
On the insident side of the boundary, the field is a composite of the
incident and reflected fields.
On the other side of the boundary the field is the refracted field.
Evaluating Eq.~(\ref{EQf1.12}) in terms of
Eqs.~(\ref{EQp1.07a})--(\ref{EQp1.07c}) 
and using the fact that the line elements $\Delta{\bf l}_1$ and
$\Delta{\bf l}_2$ in Eq.~(\ref{EQf1.12}) are equal and opposite,
we obtain a relation
\begin{equation}
\frac{{n_e}_1}{{n_m}_1} (\tilde A_i-\tilde A_r)=
\frac{{n_e}_2}{{n_m}_2}\tilde A_R
\label{EQf1.13}
\end{equation}
between the amplitudes of the incident, reflected, and refracted
vector potentials.
This relation is usually derived in the Maxwell--Minkowski
representation by continuity of the parallel ${\bf H}$ field,
{\it cf.,} Eq.~8.107 of Griffiths. \cite{BIGriffiths}
\par
In order to derive boundary relations, we need two linearly
independent first-order relations.
Substituting Eq.~(\ref{EQf1.13}) into Eq.~(\ref{EQp1.11}), we have
a unique decomposition of Eq.~(\ref{EQp1.11})
\begin{equation}
\frac{{n_e}_1}{{n_m}_1} \left (\tilde A_i - \tilde A_r \right ) =
\frac{{n_e}_2}{{n_m}_2} \tilde A_R
\label{EQf1.14}
\end{equation}
\begin{equation}
\tilde A_i + \tilde A_r = \tilde A_R 
\label{EQf1.15}
\end{equation}
that corresponds to the usual electromagnetic boundary conditions
\cite{BIMarion,BIGriffiths,BIJack,BIZang}.
\par
We eliminate $\tilde A_R$ from Eq.~(\ref{EQf1.14}) using
Eq.~(\ref{EQf1.15}) to obtain
\begin{equation}
\frac{\tilde A_r}{\tilde A_i} =
\frac{{n_e}_1{n_m}_2-{n_e}_2{n_m}_1}{{n_e}_1{n_m}_2+{n_e}_2{n_m}_1} \,.
\label{EQf1.16}
\end{equation}
Subsequently, we eliminate $\tilde A_r$ to get
\begin{equation}
\frac{\tilde A_R}{\tilde A_i}=\frac{2{n_e}_1{n_m}_2}
{{n_e}_1{n_m}_2+{n_e}_2{n_m}_1} \,.
\label{EQf1.17}
\end{equation}
Eqs.~(\ref{EQf1.16}) and (\ref{EQf1.17}) are the usual Fresnel relations
for fields normally incident on a transparent magneto-dielectric
material.
Although the formulas, Eqs.~(\ref{EQf1.16}) and (\ref{EQf1.17}), are not
novel, their derivation from conservation of energy and the wave
equation is more general than the usual derivation based on a
Minkowski representation of the macroscopic Maxwell equations
\cite{BIMarion,BIGriffiths,BIJack,BIZang}.
\par
Conservation of energy, by itself, is sufficient to derive
Eq.~(\ref{EQp1.11}).
However, there are several ways that Eq.~(\ref{EQp1.11}) can be
decomposed into two first-order equations.
The application of Stokes's theorem to the wave equation guarantees
uniqueness of the decomposition of the energy balance,
Eq.~(\ref{EQp1.11}), into boundary conditions, Eqs.~(\ref{EQf1.14})
and (\ref{EQf1.15}), and the Fresnel relations, Eqs.~(\ref{EQf1.16})
and (\ref{EQf1.17}), for stationary, transparent, linear,
magneto-dielectric, optical materials.
\par
For oblique incidence at an angle $\theta_i$ with the normal to the
interface, the width of the refracted pulse is
$w_i({n}_1 \cos\theta_R)/({n}_2\cos\theta_i)$.
The energy balance, Eq.~(\ref{EQp1.10}) becomes
\begin{equation}
\frac{{n_e}_1}{{n_m}_1} \tilde A_i^2w_i
=\frac{{n_e}_1}{{n_m}_1}\tilde A_r^2w_i
+\frac{{n_e}_2}{{n_m}_2} \tilde A_R^2w_i
\frac{\cos\theta_R}{\cos\theta_i}
\label{EQf1.18}
\end{equation}
for our nominally square/rectangular pulse.
Then the usual Fresnel relations
\begin{equation}
\frac{\tilde A_r}{\tilde A_i}=\frac
{{n_e}_1{n_m}_2 \cos\theta_i-{n_e}_2{n_m}_1\cos\theta_R}
{{n_e}_1{n_m}_2 \cos\theta_i+{n_e}_2{n_m}_1\cos\theta_R}
\label{EQf1.19a}
\end{equation}
\begin{equation}
\frac{\tilde A_R}{\tilde A_i}=\frac {2{n_e}_1{n_m}_2 cos\theta_i}
{{n_e}_1{n_m}_2 \cos\theta_i+{n_e}_2{n_m}_1\cos\theta_R}
\label{EQf1.19b}
\end{equation}
are obtained when we combine the energy balance, Eq.~(\ref{EQf1.18}),
with the relation
\begin{equation}
\tilde A_i+\tilde A_r=\tilde A_R
\label{EQf1.20}
\end{equation}
that is obtained by applying the Stokes's theorem to the wave equation
for the case in which the vector potential is perpendicular to the plane
of incidence.
Likewise, one obtains
\begin{equation}
\frac{\tilde A_r}{\tilde A_i}=\frac
{{n_e}_2{n_m}_1 \cos\theta_i-{n_e}_1{n_m}_2\cos\theta_R}
{{n_e}_2{n_m}_1 \cos\theta_i+{n_e}_1{n_m}_2\cos\theta_R}
\label{EQf1.21a}
\end{equation}
\begin{equation}
\frac{\tilde A_R}{\tilde A_i}=\frac {2{n_e}_1{n_m}_2\cos\theta_i}
{{n_e}_2{n_m}_1 \cos\theta_i+{n_e}_1{n_m}_2\cos\theta_R}
\label{EQf1.21b}
\end{equation}
by combining the energy balance, Eq.~(\ref{EQf1.18}), with
\begin{equation}
\cos\theta_i(\tilde A_i-\tilde A_r)=\cos\theta_R\tilde A_R \, .
\label{EQf1.22}
\end{equation}
that is derived from the wave equation by using the Stokes's theorem in
the case of a vector potential that is parallel to the plane of
incidence.
\par
In this communication, we obtained the electromagnetic boundary conditions
and the Fresnel relations using a derivation that has a well-defined
condition of applicability in terms of energy conservation and the validity
of the wave equation.
The new derivation of the boundary conditions and Fresnel relations is
necessary and valuable because the usual derivation is inconsistent with the
field equations in some representations of the macroscopic Maxwell field
equations for simple linear materials.
\par

\end{document}